\documentclass{PoS}

\usepackage{amsmath}
\usepackage{amssymb}
\usepackage{mathrsfs}
\usepackage{amsfonts}
\usepackage{bbm}
\usepackage{bbold}

\title{Topology of Trace Deformed Yang-Mills Theory}

\ShortTitle{Topology and Trace Deformation}

\author{\speaker{Marco Cardinali}\\
        University of Pisa and INFN-Pisa\\
        E-mail: \email{marco.cardinali@pi.infn.it}}

\author{Claudio Bonati\\
        University of Pisa and INFN-Pisa\\
        E-mail: \email{claudio.bonati@unipi.it}}

\author{Massimo D'Elia\\
        University of Pisa and INFN-Pisa\\
        E-mail: \email{massimo.delia@unipi.it}}

\author{Fabrizio Mazziotti\\
        University of Pisa and INFN-Pisa\\
        E-mail: \email{fabrizio.mazziotti@pi.infn.it}}

\abstract{In this paper we study, by means of numerical simulations,
the topological properties of $SU(3)$ and $SU(4)$ trace deformed Yang-Mills 
theory defined on $ \mathbb{R}^3\times S^1$, in which center symmetry is recovered
even at small compactification radii. In particular, we compute the topological
suscpetibility $\chi$ and the coefficient $b_2$  (related to the fourth cumulant of the topological
charge distribution). We find that these observables computed
in the deformed theory when center symmetry is recovered are compatible with their
values at zero temperature both for 3 and 4 colours.}

\FullConference{37th International Symposium on Lattice Field Theory - Lattice2019\\
		16-22 June 2019\\
		Wuhan, China}

\begin{document}

\section{Introduction}
Pure gauge Yang-Mills theories, defined on a space-time with
one or more compactified directions and periodic boundary
conditions, are invariant under center symmetry transformations. 
A center symmetry transformation consists of a gauge 
transformation which respects the periodicity but for a global element
of the center of the gauge group, in particular the center of $SU(N)$ is $\mathbb{Z}_N$.
In the limit of small compactification radius, the system undergoes a 
phase transition and center symmetry is spontaneously broken; the order
parameter of this phase transition is the holonomy around the compactified
direction and it is called Polyakov loop. If we identify the compactified direction
with the thermal Euclidean direction the spontaneous symmetry breaking
of center symmetry is the well known deconfinement phase transition and the
Polyakov loop is defined as  
\begin{equation}
P(\vec{x}) = \mathcal{P}\exp\left(i\int_0^{L}A_0(\vec{x}, \tau)
\mathrm{d}\tau\right) \,.
\label{poldef}
\end{equation}
In the low-$T$ confined phase the trace of the Polyakov loop is zero, while it becomes 
non-zero in the high-$T$ (or small compactification length $L$) regime.
The low-$T$ phase possesses several interesting properties such as confinement
or the $\theta$ dependence; however, this regime is characterized by strong coupling and
a perturbative description is not reliable. On the contrary, the high-$T$ regime is weakly 
coupled and it is possible, for example, to use semiclassical approximations to make predictions.
Since the low-$T$ and high-$T$ phases are separeted by a phase transition they are not
analitically connected and it is not possible to use the semiclassical methods 
mentioned before to describe the confining phase.\\
Trace deformed Yang-Mills was first studied on the lattice in~\cite{Myers:2007vc} and then
proposed as a possible way to overcome the problem of the phase transition in~\cite{Unsal:2008ch}. 
The deformation consists of extra pieces added to the usual 
Yang-Mills action which preserve center
symmetry even at high-$T$. The discretized action of the theory for $SU(3)$ is the following:

\begin{equation}
  S^{\mathrm{def}} = S_{YM} + h \sum_{\vec n} |\mathrm{Tr} P(\vec n)|^2 
\label{tracedef_su3}
\end{equation}
where $h$ is the coupling of the trace deformation and $S_{YM}$ is the usual
Wilson action. If the parameter $h$ is high enough, the extra piece
suppresses gauge configurations in which the Polyakov loop acquires a non zero value;
thus the system recover center symmetry even at high temperature.\\
In this paper we compute on the lattice two observables related to the topological properties 
of the deformed theory when center symmetry is recovered: the topological susceptibility $\chi$ and the coefficient $b_2$, 
which is related to the fourth cumulant of the topological charge distribution. A 
first study (see Ref.~\cite{gufo}) showed that, for $N=3$, both the topological susceptibility and 
$b_2$ reach a plateau value increasing the parameter $h$, and such values are compatible with
the zero temperature ones. Inspired by this result we decided to extend the analysis
to $N=4$ (see Ref.~\cite{gufo2}), since in this case we have a less trivial center symmetry breaking
pattern. In section 2 we will give a brief review of $\theta$ 
dependence and how to determine it on the lattice. Section 3 is dedicated
to $SU(3)$ results, while section 4 to $SU(4)$. In the end we will summarize the main results
in the conclusions.

\section{$\theta$ Dependence and Lattice Setup}
The dependence on the topological $\theta$-parameter enters 
in the (Euclidean) Lagrangian as follows:

\begin{equation}\label{lagrangian}
\mathcal{L}_\theta  = \frac{1}{4} F_{\mu\nu}^a(x)F_{\mu\nu}^a(x)
- i \theta q(x)\ ,
\end{equation}
where $q(x)$ is the topological charge defined by
\begin{equation}\label{topchden}
q(x)=\frac{g^2}{64\pi^2} 
\epsilon_{\mu\nu\rho\sigma} F_{\mu\nu}^a(x) F_{\rho\sigma}^a(x).
\end{equation}
The relevant information about $\theta$ dependence is contained in the
free energy density, which around $\theta = 0$ can be parametrized as follows~\cite{Vicari:2008jw}:

\begin{equation}\label{eq:theta_dep}
f(\theta) = f(0) + \frac{1}{2}\chi\theta^2(1+b_2\theta^2+b_4\theta^4+\cdots)
\end{equation}
where the topological suscpetibility $\chi$ and all the coefficients $b_{2n}$
can be related to the cumulants of the topological charge distribution computed
at $\theta=0$ by the relations:

\begin{equation}\label{eq:chi_b2}
\chi = \frac{\langle Q^2 \rangle_{c,\theta=0}}{\mathcal{V}}\ ,\quad
b_{2n}=
(-1)^n\,
\frac{2\, \langle Q^{2n+2}\rangle_{c,\theta=0}}{(2n + 2)! \langle Q^2\rangle_{c,\theta=0}}\ ,
\end{equation}
in which $Q$ is the winding number defined as $Q = \int d^4xq(x)$ and $\mathcal{V}$ is the 
four-dimensional volume.\\
Topological susceptibility is predicted to stay finite in the large-$N$ limit~\cite{Hooft-74, Witten-80, Witten-98} and in the 
low temperature, confined phase, while the other coefficients $b_{2n}$ are suppressed as follows:

\begin{equation}
\chi = \chi_\infty + O(N^{-2}), \quad  b_{2j}=O(N^{-2j}) \,.
\label{lnasyt0}
\end{equation}
Lattice simulations support such predictions, see Refs.~\cite{Lucini:2001ej,DelDebbio:2002xa,LTW-05,Ce:2016awn} for the 
topological susceptibility and Refs.~\cite{DelDebbio:2002xa,DElia:2003zne, Giusti:2007tu, Panagopoulos:2011rb, Ce:2015qha, Bonati:2015sqt, Bonati:2016tvi} for the coefficient $b_2$.\\ 
On the contrary, in the high temperature deconfined regime the theory is weakly coupled and 
one expects that semiclassical arguments, such as instanton calculus, can
be used to describe topology. In particular, the dilute instanton
gas approximation (DIGA) leads to the following expressions:

\begin{eqnarray}
f(\theta) - f(0) &\simeq& \chi(T) \left( 1 - \cos\theta\right) \nonumber \\
\chi(T) &\simeq&  T^4 \exp[-8\pi^2/g^2(T)] \sim  T^{-\frac{11}{3} N + 4},
\end{eqnarray}
which predicts, in the large-$N$
limit, a constant value for the coefficients $b_{2n}$ and a vanishing value
for the topological susceptibility. Lattice simulations~\cite{LTW-05, Bonati:2015sqt, susc_ft, GHS-02, DPV-04} performed above the
deconfinement phase transistion are in agreement with the DIGA predictions; 
moreover, the transition between the low-$T$ and high-$T$ behaviour of the
topology is sharper and sharper as $N$ increases.\\
In this paper we measure both $\chi$ and $b_2$ in the deformed theory for $SU(3)$ and 
$SU(4)$. We  performed simulations using the deformed action (Wilson action plus 
deformations) at $\theta = 0$ and then we 
computed $\chi$ on the gauge configurations 
using the standard gluonic definition, while the coefficient $b_2$ has 
been measured using the imaginary $\theta$ method exposed in~\cite{Bonati:2015sqt}. 

\section{$SU(3)$ Results}

\begin{figure}[!ht]
   \begin{minipage}{0.48\textwidth}
   \centering
   \includegraphics[width=0.9\columnwidth, clip]{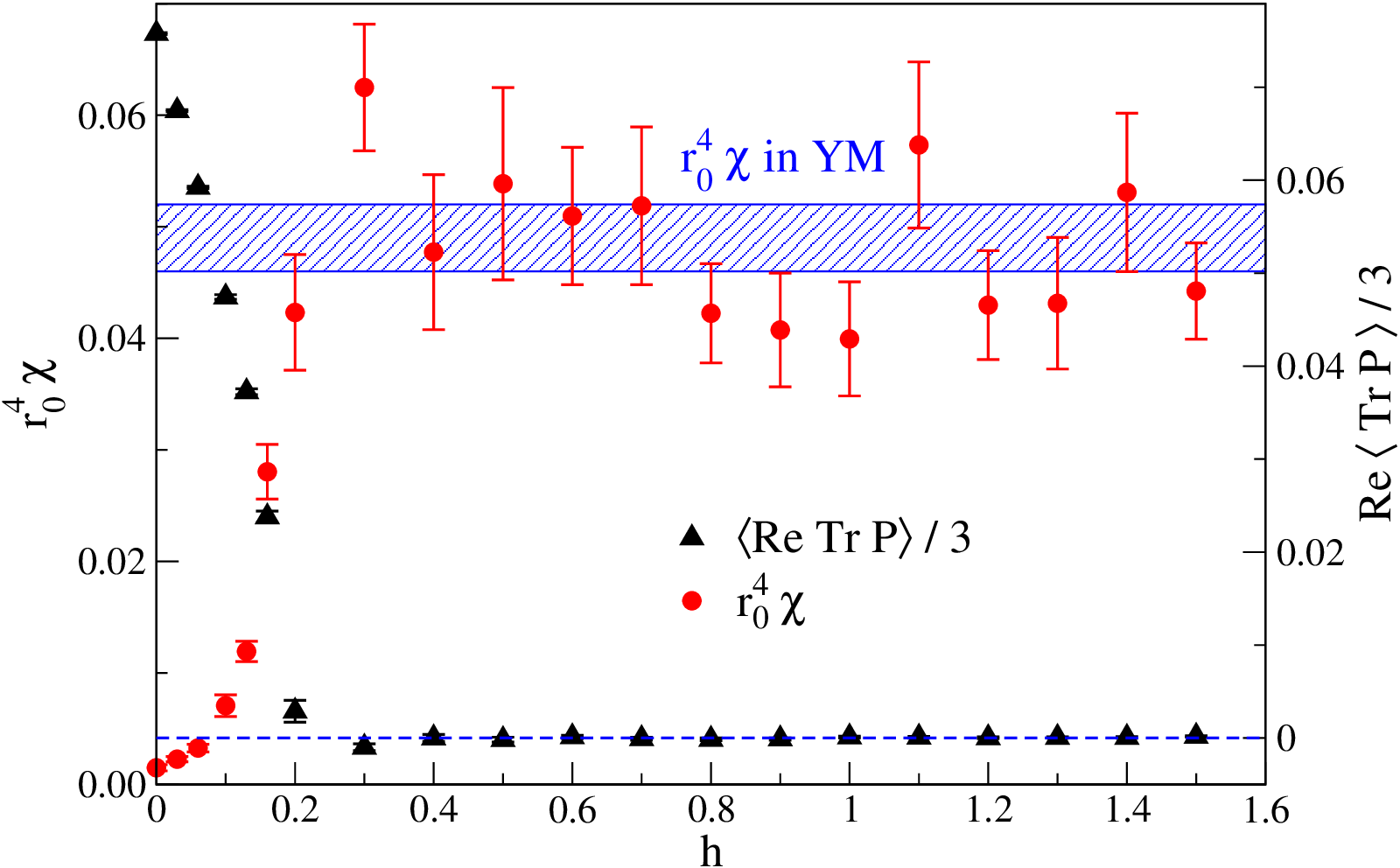}
   \caption{Topological susceptibility $\chi$  and the Polyakov loop $\mathrm{Re}\langle \mathrm{Tr}
  P\rangle/3$ as functions of the parameter $h$. The lattice is $8\times 32^3$
   and the bare coupling is $\beta=6.4$. We also report for reference
   the standart $SU(3)$ value at zero temperature~\cite{Vicari:2008jw}.}
   \label{fig:r04chi_vs_h}
   \end{minipage}\hfill
   \begin{minipage}{0.48\textwidth}
     \centering
  \includegraphics[width=0.9\columnwidth, clip]{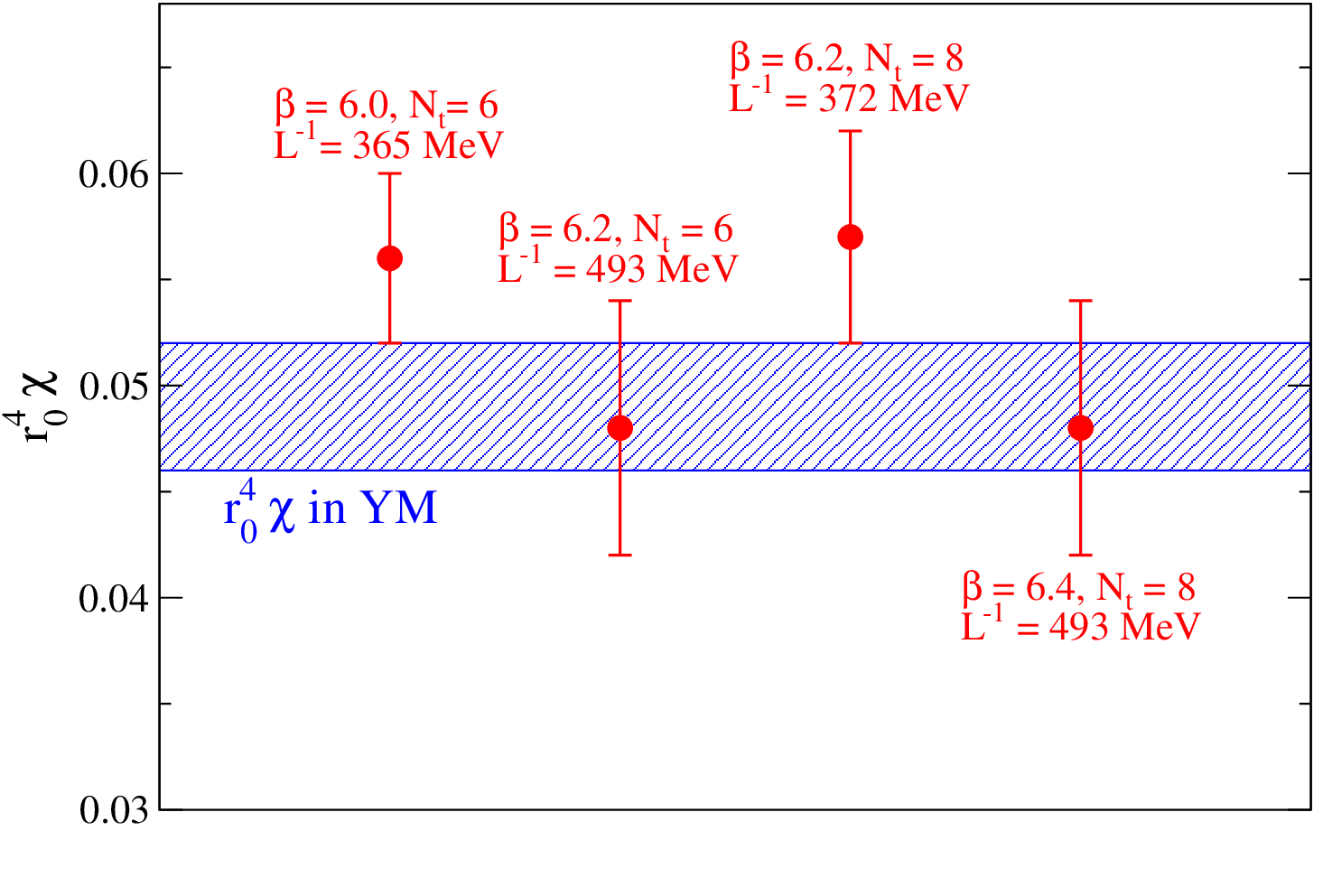}
  \caption{The plaetau value of the topological susceptibility for different temporal
         extensions and bare couplings. We also report for reference
         the standart $SU(3)$ value at zero temperature~\cite{Vicari:2008jw}.}
  \label{fig:global_r04chi_vs_h}
  \end{minipage}
\end{figure}
\begin{figure}[!ht]
   \begin{minipage}{0.48\textwidth}
   \centering
   \includegraphics[width=0.9\columnwidth, clip]{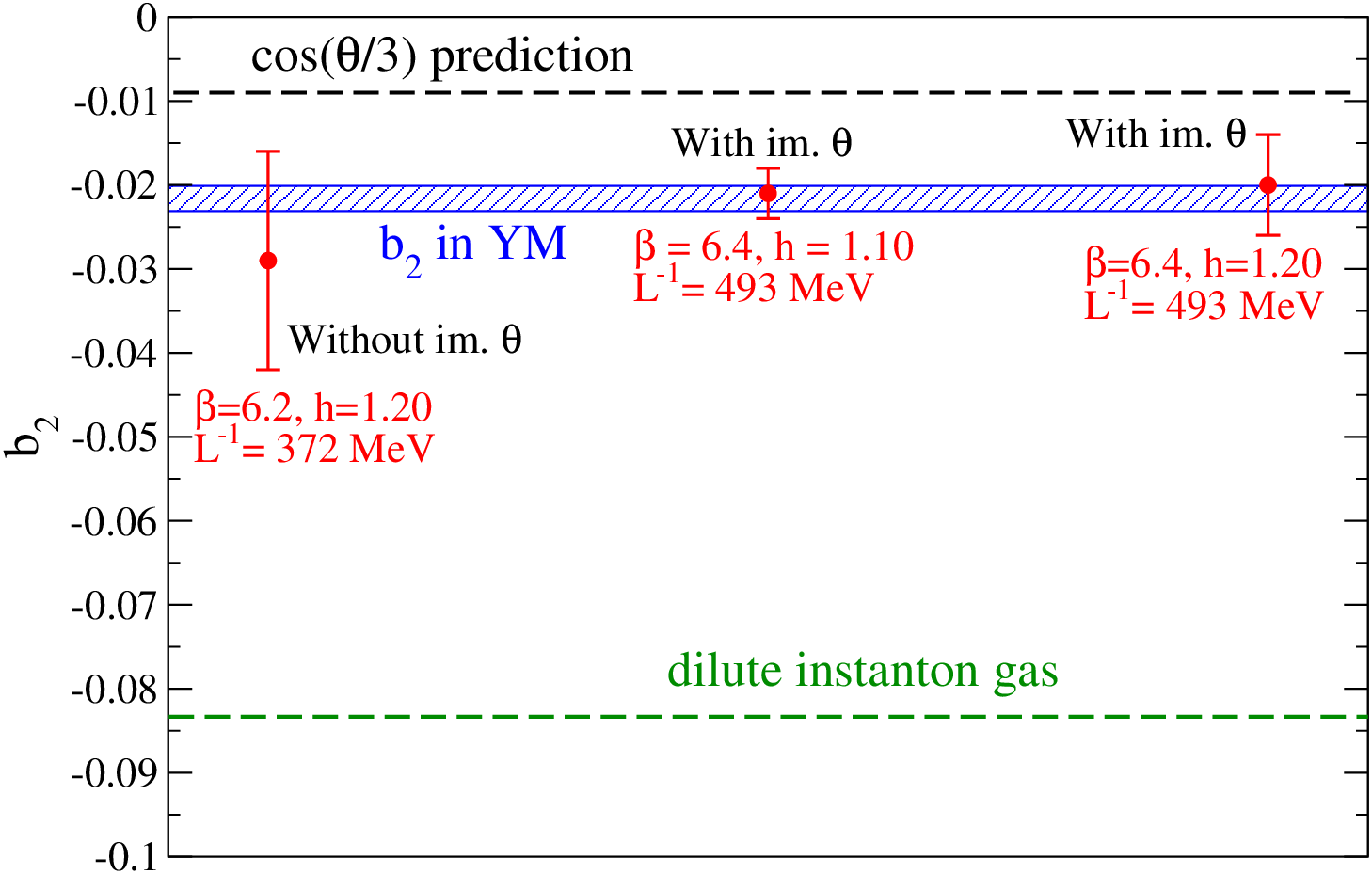}
   \caption{The coefficient $b_2$ measured using $8\times 32^3$
   lattices. The standard $SU(3)$ value
   \cite{Bonati:2015sqt} is the horizontal band, dashed lines represent the 
   value $b_2=-1/12$ (DIGA) 
   and $b_2= - 1/108$ (Fractional Instanton Gas Approximation~\cite{Thomas:2011ee,unsal-1,Aitken:2018mbb}).}
   \label{fig:b2}
  \end{minipage}\hfill
   \begin{minipage}{0.48\textwidth}
     \centering
  \includegraphics[width=0.9\columnwidth, clip]{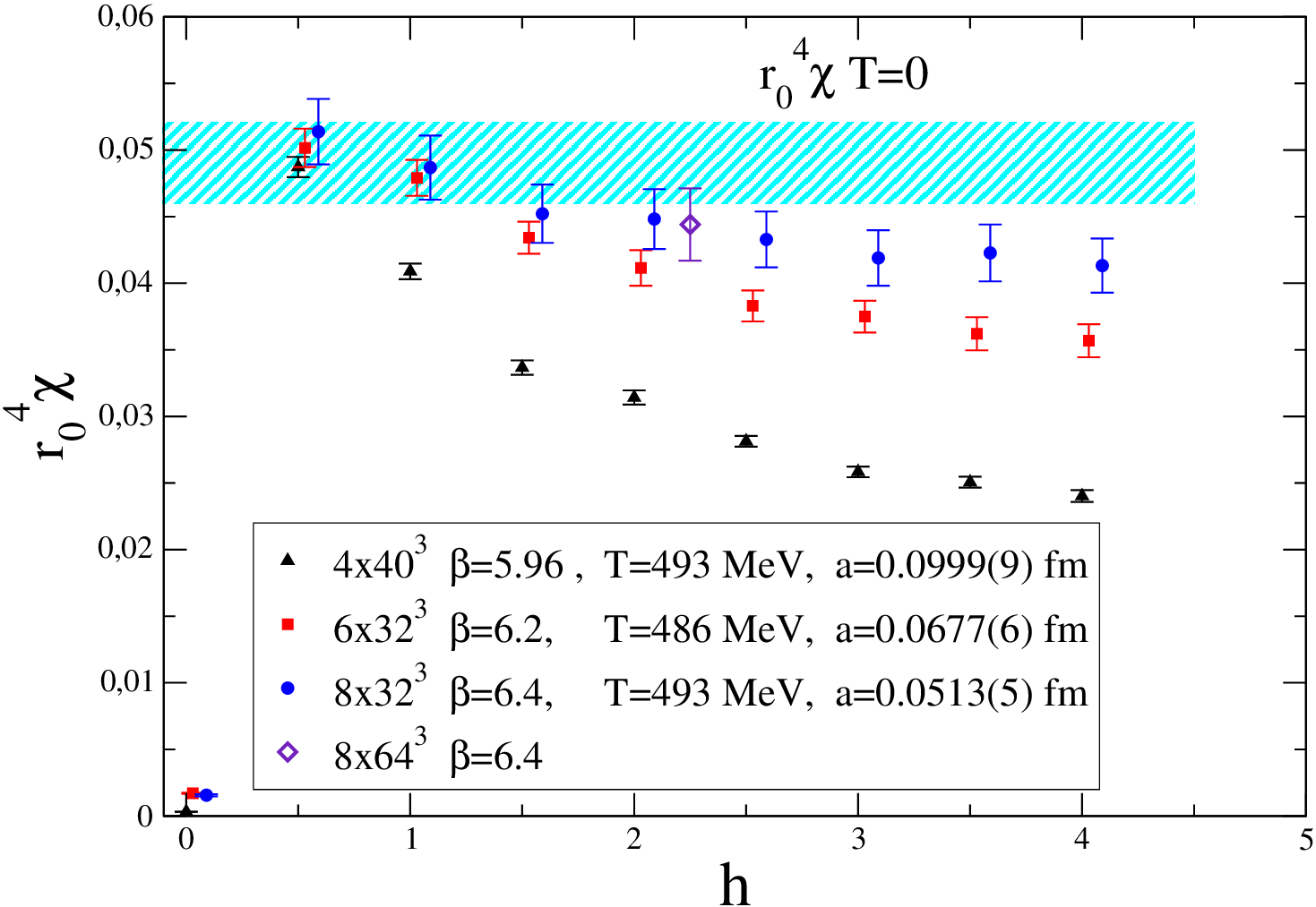}
  \caption{Topological susceptibility $\chi$ computed using different values of
           the coupling $h$ and using different lattice setups. The lattice spacing has
           been computed using $r_0=0.5$.}
         \label{fig:cut_off_effect}
  \end{minipage}
\end{figure}

In Fig.~\ref{fig:r04chi_vs_h} the topological susceptibility is shown as a function of $h$. 
The system is deep in the deconfined phase and when the deformation is off ($h=0$) 
$\chi$ is very small. As $h$ increases the topological susceptibility starts growing and reaches a plateau. The 
plateau is reached when $\mathrm{Re}\langle \mathrm{Tr} P\rangle/3 $ becomes zero, thus when
center symmetry is recovered; the remarkable fact is that the plateau value is in agreement with
the zero temperature one. In order to fix the scale we used the Sommer parameter $r_0$ fixing it at $r_0=0.5$
(see Ref.~\cite{Sommer:1993ce}) and we assumed that the deformation does not modify the lattice spacing. This has been checked
explicitly in Ref.~\cite{gufo}. We performed the same measures also with 
different bare couplings and temporal extensions and the results are shown in Fig.~\ref{fig:global_r04chi_vs_h}; 
in all the cases the plateau values are in agreement with the zero temperature result. 
We show in Fig.~\ref{fig:b2} the values of the coefficient $b_2$ in the deformed theory for two values of $h$ in the plateau region. 
Even in this case the results are compatible with the zero temperature case, while they are different
from the semiclassical high-$T$ predictions.\\
We also measured the topological susceptibility using higher values of the coupling $h$ and 
different lattice setups in order to check if there are any cut-off effects. We consider
lattices with approximately the same temperature, but with finer lattice spacings. 
Results are shown in Fig.~\ref{fig:cut_off_effect}. 
From the figure we can understand that the plateau region is more stable when
the lattice spacing is finer, i.e.\ when the continuum limit is approached.  In
particular we can see that the plateau region is not clearly defined for small $N_t$; a fact that
suggests that the deformation may lead to severe cut-off
effects when the temporal extension is too small.\\
We also checked possible finite-volume effects; we measured the topological
susceptibility using a $8\times64^3$ lattice at $h=2.0$, $\beta = 6.4$ and we
compared it with the same value obtained on a $8\times32^3$ lattice. As we
can see from Fig.~\ref{fig:cut_off_effect}, there are no visibile
finite-volume effects.

\section{$SU(4)$ Results}
$SU(4)$ Yang-Mills theory has two possible ways to break center symmetry; either the symmetry
is fully broken ($\mathbb{Z}_4 \rightarrow \mathbb{1}$) or there is a residual symmetry group left
unbroken ($\mathbb{Z}_4 \rightarrow \mathbb{Z}_2$). Therefore, in order to recover the full center symmetry we need 
two deformations: one proportional to $|\mathrm{Tr}P|$ and one proportional to $|\mathrm{Tr}P^2|$, where the second deformation is used to avoid the breaking to $\mathbb{Z}_2$ since the order
parameter of this phase transition is $\mathrm{Tr}P^2$. The action used in the simulation is the following:

\begin{equation}
  S^{\mathrm{def}} = S_{YM} + h_1 \sum_{\vec n} |\mathrm{Tr} P(\vec n)|^2 
  + h_2 \sum_{\vec n} |\mathrm{Tr} P^2(\vec n)|^2. 
\label{tracedef_su4}
\end{equation}

\begin{figure}[!ht]
   \begin{minipage}{0.48\textwidth}
   \centering
   \includegraphics[width=0.9\columnwidth, clip]{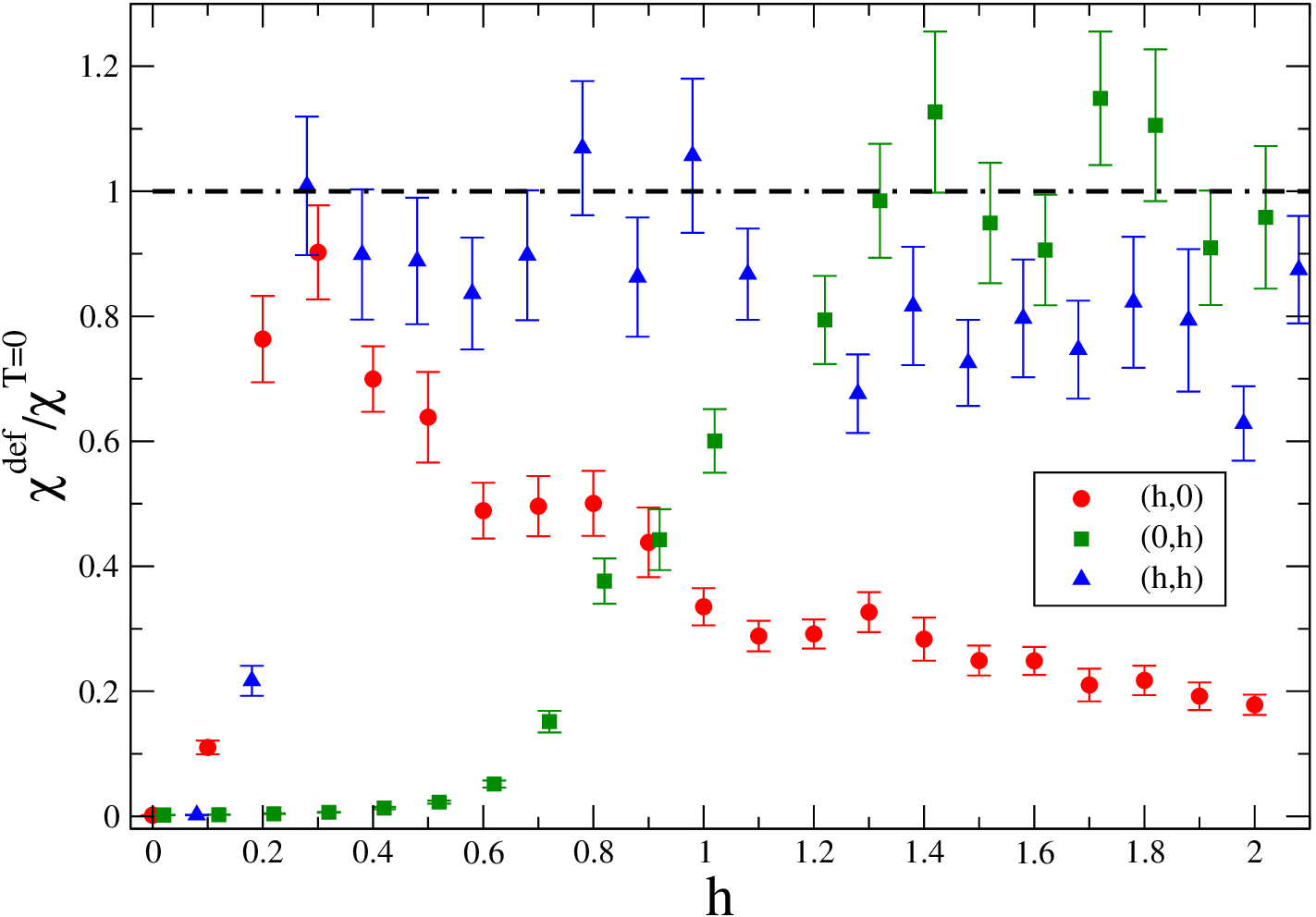}
   \caption{Topological susceptibility $\chi$ as a function of the 
     parameters $h_1$ and $h_2$. The lattice is $6\times 32^3$
   and the bare coupling is $\beta=11.40$. }
   \label{fig:chi_su4}
   \end{minipage}\hfill
   \begin{minipage}{0.48\textwidth}
     \centering
  \includegraphics[width=0.9\columnwidth, clip]{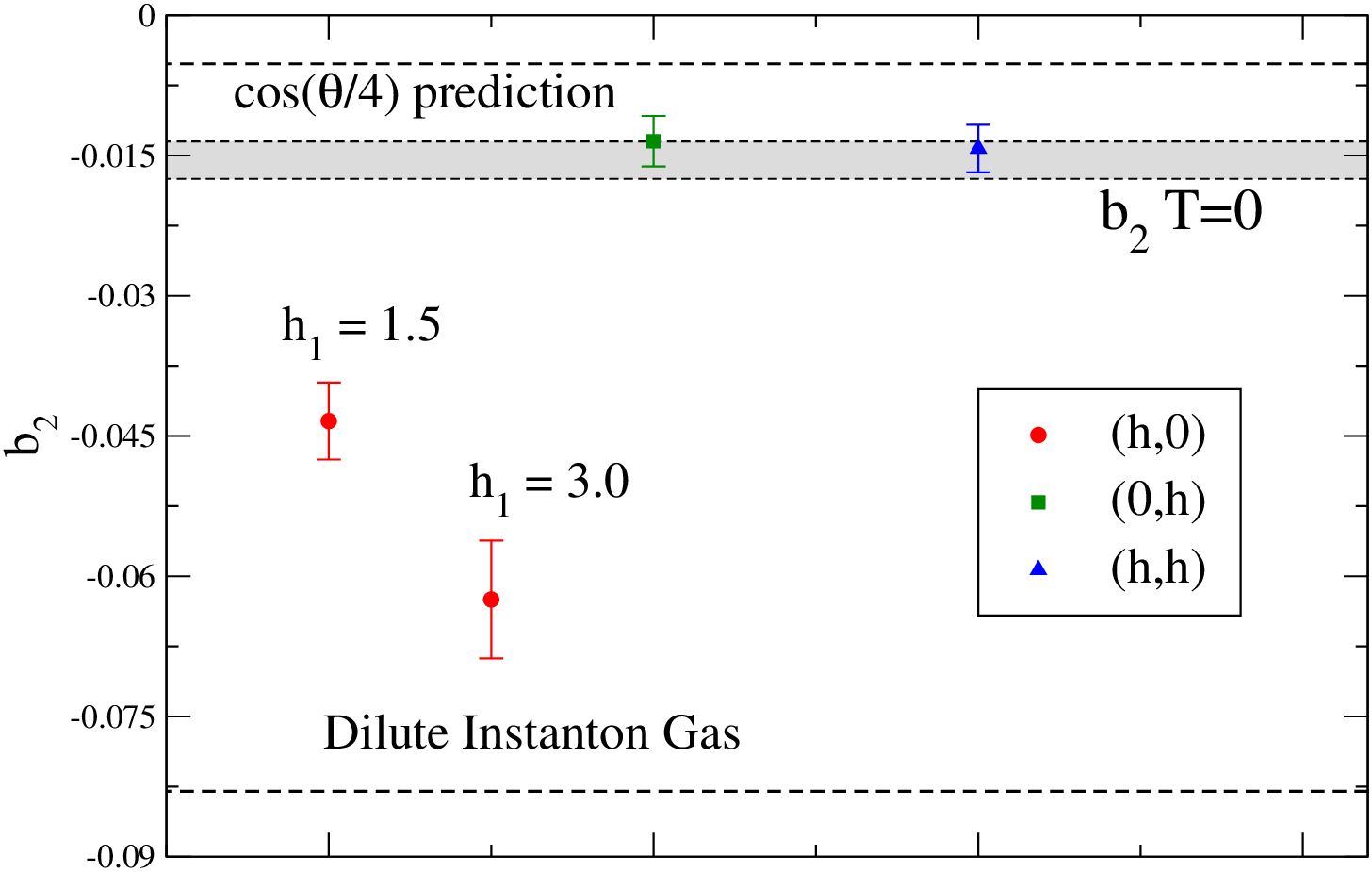}
  \caption{coefficient $b_2$ measured in the deformed theory on a $6\times32^3$ lattice
  at $\beta=11.40$ using one or both deformations; $h_1$ and $h_2$ have been
chosen in the plateau region of the topological susceptibility.}
       \label{fig:b2_su4}
  \end{minipage}
\end{figure}

In Figs.~\ref{fig:chi_su4} and~\ref{fig:b2_su4} we show the results for $SU(4)$. If we use both
deformations (triangular points) the topological susceptibility and $b_2$ become compatible with the $T=0$ values 
while if we use only one of the deformation the situation is different. The term proportional to $|\mathrm{Tr}P|$ is 
not able to recover full center symmetry and both the topological suscpetibility and $b_2$ do 
not reach a plateau (circular points). However, if we use only the term proportional to
$|\mathrm{Tr}P^2|$ we are able to recover full center symmetry and $\chi$ and $b_2$ behave as
in the zero temperature case.

\section{Conclusions}
We studied the $\theta$-dependence of the trace deformed Yang-Mills theory using
3 and 4 colors. The trace deformation consists of additional terms in the standard Yang-Mills
action used to prevent the spontaneous symmetry breaking of center symmetry even at small compactification
radii. We computed, by means of numerical simulations, the topological susceptibility $\chi$ and 
the coefficient $b_2$ related to the fourth order cumulant of the expansion of the topological 
charge distribution. Both in $SU(3)$ and $SU(4)$ these observables computed in the deformed
theory are compatible with their values obtained at zero temperature; however, for $SU(4)$ 
we observed that center symmetry must be fully recovered in order to obtain the $T=0$ results.

\end{document}